\documentclass{INTERSPEECH2023}

\interspeechcameraready

\usepackage[textsize=footnotesize]{todonotes}
\usepackage{tabularray}
\SetTblrInner{rowsep=0pt} 
\usepackage{svg}
\usepackage{xfrac}
\usepackage{balance}
\usepackage{amsmath}
\usepackage{comment}

\newcommand{\signal}{\mathbf{s}}

\newcommand{\xvdim}{m}

\newcommand{\xvo}{\mathbf{x}_o}
\newcommand{\XVo}{\mathbf{X}_o}

\newcommand{\xvp}{\mathbf{x}_p}
\newcommand{\XVp}{\mathbf{X}_p}

\newcommand{\xva}{\mathbf{x}_a}

\newcommand{\xvi}{\mathbf{x}_i}
\newcommand{\XVi}{\mathbf{X}_i}

\newcommand{\fno}{\mathbf{f}}
\newcommand{\ppg}{\mathbf{G}}

\DeclareMathOperator{\vcdr}{V}
\newcommand{\vocoder}[3]{\vcdr\left(#1, #2, #3 \right)}

\newcommand{\anon}[1]{a\left(#1\right)}

\newcommand{\secref}[1]{Section~\ref{#1}}
\newcommand{\figref}[1]{Figure~\ref{#1}}
\newcommand{\tabref}[1]{Table~\ref{#1}}

\DeclareMathOperator*{\argmin}{argmin}

\title{Vocoder drift compensation by x-vector alignment in speaker anonymisation}
\name{Michele Panariello, Massimiliano Todisco, Nicholas Evans}
\address{EURECOM, Sophia Antipolis, France}
\email{firstname.lastname@eurecom.fr}

\begin{document}

\maketitle
 
\begin{abstract}
For the most popular x-vector--based approaches to speaker anonymisation, the bulk of the anonymisation can stem from vocoding rather than from the core anonymisation function which is used to substitute an original speaker x-vector with that of a fictitious pseudo-speaker.  This phenomenon can impede the design of better anonymisation systems since there is a lack of fine-grained control over the x-vector space.  The work reported in this paper explores the origin of so-called vocoder drift and shows that it is due to the mismatch between the substituted x-vector and the original representations of the linguistic content, intonation and prosody.  Also reported is an original approach to vocoder drift compensation.  While anonymisation performance degrades as expected, compensation reduces vocoder drift substantially, offers improved control over the x-vector space and lays a foundation for the design of better anonymisation functions in the future.
\end{abstract}
\noindent\textbf{Index Terms}: anonymisation, pseudonymisation, privacy, vocoder drift, automatic speaker verification

\section{Introduction}
\label{sec:intro}
\emph{Speaker anonymisation} broadly refers to the task of processing speech recordings to conceal the identity of the speaker while preserving linguistic and paralinguistic content.
Recently, the topic has attracted notable research interest, particularly through the VoicePrivacy Challenge~\cite{introducing_vp, vp2020results, vp2022eval}, first launched in 2020 to define the task and to encourage the development of more effective speaker anonymisation techniques.
According to the VoicePrivacy Challenge Evaluation Plan~\cite{vp2022eval}, the evaluation of a speaker anonymisation solution is based upon estimates of the trade off between \emph{privacy} (protection of the speaker identity) and \emph{utility} (how well the remaining signal content is preserved). The former is estimated by the ability of an attacker to use automatic speaker verification (ASV) to infer the original speaker identity and is measured in terms of equal error rate (EER).  The latter is estimated using the word error rate (WER) of an automatic speech recognition (ASR) system as a proxy for utility.

Currently, the better-performing anonymisation solutions reflect the processing pipeline described in~\cite{Fang2019} and rely upon an initial decomposition of the input signal into the following three components:
\begin{itemize}
    \item a set of features representing the linguistic content of the signal, typically in the form of ASR;
    \item a component representing intonation and prosody, normally a fundamental frequency (F0) curve;
    \item a neural embedding representing the identity of the speaker, usually an \emph{x-vector}.
\end{itemize}
To obfuscate the speaker identity, an \emph{anonymisation function} is applied to the x-vector embedding, thereby obtaining a new embedding which represents the voice of a fictitious \emph{pseudo-speaker}. The three components are subsequently fed to a \emph{vocoder} model which synthesises a waveform with the same spoken content and prosody as the original input audio, but in the voice of the substitute pseudo-speaker.

For effective anonymisation, the pseudo-speaker's voice should sound \emph{different} to that of the original speaker. 
For the majority of anonymisation systems proposed to date, this requirement is fulfilled by maximising some measure of the distance between the chosen pseudo-speaker embedding and the original speaker embedding.
The most popular anonymisation function to date uses a \emph{pool} of external \mbox{x-vectors}~\mbox{\cite{vp2020results, vp2022eval, Fang2019, champion_f0, vp_T40, miao22_odyssey}}. The pseudo-speaker embedding is obtained by averaging a random subset of the furthest x-vectors in the pool from the x-vector of the original speaker. More refined methods of pseudo-speaker selection, e.g. based upon the use of singular-value decomposition~\cite{anon_svd} and generative adversarial networks~\cite{anon_gan}, have also been explored.  
However,
in our previous work~\cite{drift}, we showed that anonymisation performance is influenced by more than just the role of the anonymisation function.
The vocoder
also plays a role and its impact 
is comparable to, or even dominates that of the anonymisation function. 
We termed this phenomenon \emph{vocoder drift}.

While one interpretation of these observations is that vocoder drift contributes positively to anonymisation, and is hence a benefit, another is that it implies a lack of fine-grained control over the x-vector space and that this lack of control in turn impedes the design of effective x-vector anonymisation functions.
Moreover, 
vocoder drift can be learned and reversed to undermine anonymisation safeguards~\cite{drift}.

With the work reported in this paper, we have sought to understand the cause of vocoder drift and how it can be reduced in order to improve 
control over the x-vector trajectory and full anonymisation process.
We show that the cause of drift is related to the mismatch between the distribution of x-vectors used for vocoder training and the distribution after anonymisation.
Such a mismatch can be compensated for during anonymisation by aligning the input and output x-vectors of the vocoder via gradient descent.

\section{Relation to prior work}
\label{sec:prior_work}
In this section, we describe the typical structure of an x-vector--based anonymisation solution, the concept of drift, and other relevant, prior work.
We also describe the system we used for the experiments reported in Sections~\ref{sec:causes}~and~\ref{sec:align}.

\subsection{X-vector--based anonymisation}
\label{sec:xv_anon}
\begin{figure}[t]
    \centering
    \includegraphics[width=\columnwidth]{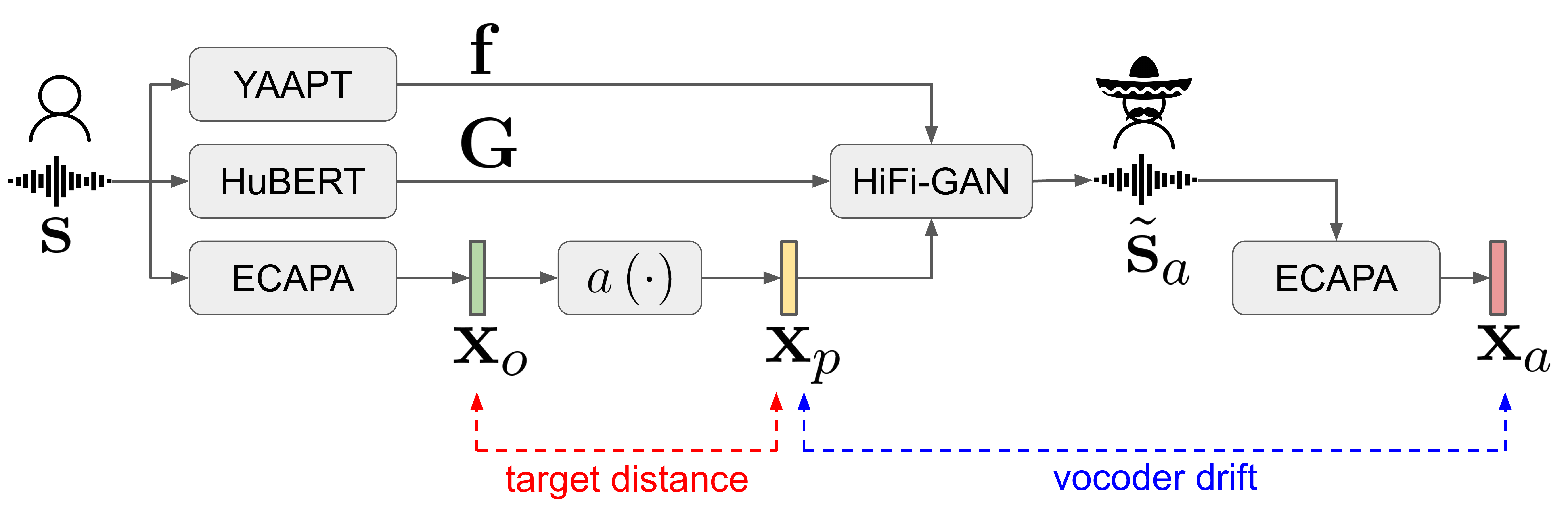}
    \caption{Diagram of the speaker anonymisation system used in this work. The dashed arrows indicate which x-vectors are used to compute \emph{target distance} and \emph{vocoder drift}.}
    \label{fig:anon_system}
\end{figure}

A graphical overview of a typical x-vector--based approach to anonymisation is shown in \figref{fig:anon_system}.
Let $\signal$ be a speech signal
which we seek
to anonymise and from which we extract the following components: an F0 curve $\fno\in\mathbb{R}^{N}$, where $N$ is the number of frames into which $\signal$ is split; a set of linguistic features $\ppg\in\mathbb{R}^{c \times N}$, where $c$ is the feature dimension; an x-vector $f(\signal) = \xvo \in \mathbb{R}^{\xvdim}$, where $m$ is the embedding dimension and $f(\cdot)$ is the embedding extraction function.
The x-vector is duplicated once for each frame, resulting in a matrix $\XVo \in \mathbb{R}^{\xvdim \times N}$.
The set of features are then concatenated into a final matrix of dimension $(1 + c + m) \times N$ and fed to a vocoder model to produce an utterance $\Tilde{\signal}=\vocoder{\fno}{\ppg}{\XVo}$.\footnote{Henceforth, a bold lowercase $\mathbf{x}$ refers to a single x-vector, while an uppercase $\mathbf{X}$ of the same subscript represents the matrix constructed from the same x-vector duplicated $N$ times.}

The vocoder is trained in a self-supervised fashion to reconstruct the original signal.
At test time, an input utterance is anonymised by substituting the original speaker embedding $\xvo$ with a pseudo-speaker embedding $\xvp$, which is obtained by means of an anonymisation function  $\anon{\xvo}=\xvp$. The anonymised utterance $\Tilde{\signal}_a$ is synthesised as $\Tilde{\signal}_a = \vocoder{\fno}{\ppg}{\XVp}$, and a further x-vector $\xva = f(\Tilde{\signal}_a)$ can then be extracted from it.
Thus, as a result of anonymisation, the speaker identity follows an x-vector trajectory, from $\xvo$ to $\xvp$ and then $\xva$.

\begin{table*}[!t]
\caption{Average target distance and drift (without and with compensation) for the four different VoicePrivacy Challenge 2022 data subsets and for four different values of $\lambda$.}
\label{tab:drift}
\centering
\resizebox{\linewidth}{!}{%
\footnotesize
\begin{tblr}{|l|ccc|ccc|ccc|ccc|}
\hline[1pt]
\SetCell[r=2]{} & \SetCell[c=3]{} $\lambda=0$ (copy-synthesis) & & & \SetCell[c=3]{} $\lambda=1/3$ & & & \SetCell[c=3]{} $\lambda=1/2$ & & & \SetCell[c=3]{} $\lambda=1$ (normal anon.) \\ 
\cline{2-13} & target & drift & drift {\scriptsize (compens.)} & target & drift & drift {\scriptsize (compens.)} & target & drift & drift {\scriptsize (compens.)} & target & drift & drift {\scriptsize (compens.)} \\ \hline 
LibriSpeech (F) & 0 & 0.29 & 0.047 & 0.13 & 0.38 & 0.049 & 0.35 & 0.50 & 0.054 & 1.0 & 0.63 & 0.052 \\
LibriSpeech (M) & 0 & 0.27 & 0.047 & 0.11 & 0.35 & 0.048 & 0.31 & 0.48 & 0.051 & 1.0 & 0.65 & 0.052 \\
VCTK (F) & 0 & 0.29 & 0.049 & 0.11 & 0.36 & 0.051 & 0.30 & 0.49 & 0.084 & 1.0 & 0.69 & 0.082 \\
VCTK (M) & 0 & 0.29 & 0.049 & 0.08 & 0.35 & 0.049 & 0.26 & 0.46 & 0.062 & 1.1 & 0.79 & 0.078 \\ \hline[1pt]
\end{tblr}}
\end{table*}

\subsection{Vocoder drift}
\label{sec:drift_def}
In~\cite{drift}, we sought to understand the degree to which the anonymisation function and the speech synthesis procedure impact upon the x-vector trajectory from $\xvo$ to $\xva$. We did so by measuring how much the x-vector is perturbed during these two steps of the anonymisation pipeline.

The anonymisation function controls the shift from $\xvo$ to~$\xvp$. Given a distance metric $d$, we define $d(\xvo, \xvp)$ as the \emph{target distance}: this quantity is set by the system designer and indicates the desired perturbation which is applied to the original speaker embedding to give the  pseudo-speaker embedding.
As a result of synthesis, $\xvp$ is further perturbed by the vocoder, giving $\xva$.  We term $d(\xvp, \xva)$ the \emph{vocoder drift}.
To provide fine-grained control over the x-vector space, the impact of drift should be as small as possible in the total trajectory of an individual x-vector. In other words, ideally, $d(\xvo, \xvp) \gg d(\xvp, \xva)$.

Our work reported in~\cite{drift} shows that the impact of the anonymisation function and vocoder are comparable and that, in some cases, the bulk of the anonymisation is delivered by the vocoder, not the anonymisation function.
In this work, we propose a technique to compensate for vocoder drift.

\subsection{System setup}
\label{sec:system_setup}
For all experiments reported in this paper, we use a system which, except for the use of different vocoders, is the same as that described in~\cite{drift}, which is itself inspired by original work in~\cite{miao22_odyssey}.
The F0 contour~$\fno$ and the linguistic features~$\ppg$ are produced with YAAPT~\cite{yaaaaaaapt} and a HuBERT-based soft content encoder~\cite{hubert}, respectively. All x-vectors are extracted with ECAPA-TDNN~\cite{ecapa}, and the vocoder model is a HiFi-GAN~\cite{hifi_gan}.
The anonymisation function $\anon{\cdot}$ is the same x-vector pool-based averaging function described in \secref{sec:intro}. Given an input $\xvo$, the $K$ x-vectors furthest from it are selected from the pool.  $K^*$ of them are then randomly chosen and averaged to obtain $\xvp$.
We set $K=200$, $K^{*}=100$, and use the cosine distance metric as in~\cite{drift}.
Following the VoicePrivacy Challenge 2022~\cite{vp2022eval} protocol, the external x-vector pool is derived from \textit{LibriTTS-train-other-500}~\cite{libriTTS}, and the evaluation sets are derived from the \textit{LibriSpeech-test-clean}~\cite{librispeech} and VCTK~\cite{vctk} (split into female and male sub-partitions) datasets.
For consistency with $\anon{\cdot}$, the target distance and vocoder drift are also measured in terms of the cosine distance.

\begin{figure}[t]
    \centering
    \includegraphics[width=\columnwidth]{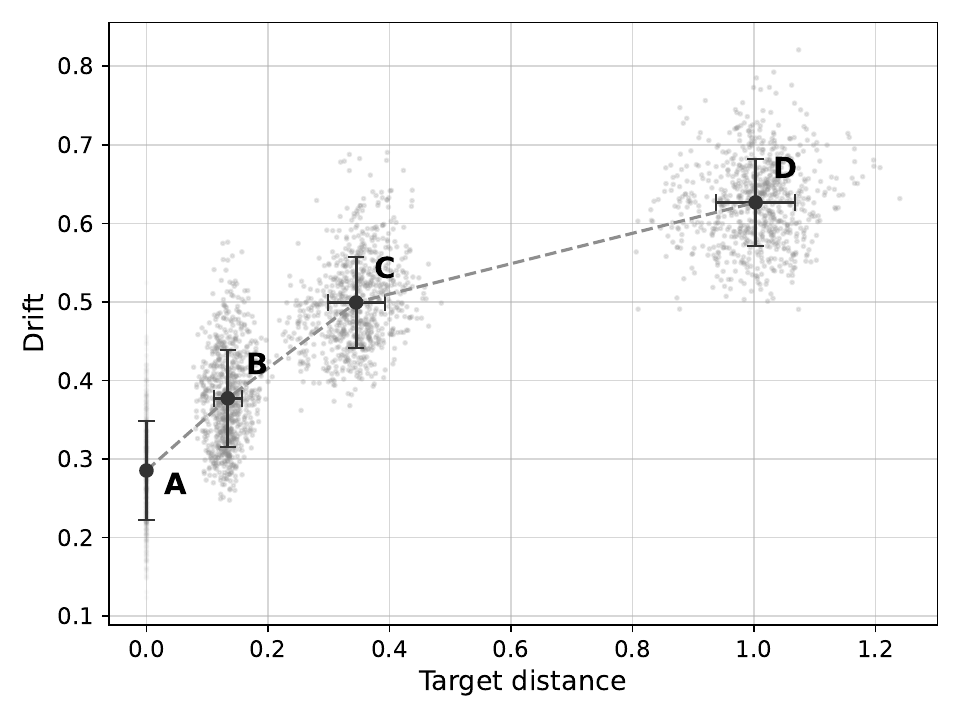}
    \caption{Vocoder drift plotted against target distance for the LibriSpeech dataset and for female speakers. Dots and bars represent mean and standard deviation of each of the following experimental setups: (A) $\lambda=0$; (B) $\lambda=1/3$; (C) $\lambda=1/2$; (D) $\lambda=1$.}
    \label{fig:drift_trend}
\end{figure}

\section{The cause of vocoder drift}
\label{sec:causes}
In this section, we describe what we believe to be the source of vocoder drift and present a set of experiments which validate our hypothesis.

\subsection{Feature mismatch}
As illustrated in \secref{sec:xv_anon}, the vocoder model is trained in self-supervised fashion to reconstruct input signals $\signal$ at the output.
While, ideally, input components $\fno$, $\ppg$ and $\xvo$ should be disentangled from one another -- so that none contains any information that is also contained in any other -- there is no explicit incentive in the training criterion of any of the three extraction models which would encourage the learning of disentangled representations.  Previous work has confirmed that the representations are indeed \emph{entangled} to some extent.  For example, results in~\cite{are_disentangled_representations, dp_spk_anon} show that speaker-related information, 
normally captured in $\xvo$,  
can leak into linguistic representations~$\ppg$.
The vocoder can hence learn to rely on such mutual dependencies between input features in learning how it should reconstruct~$\Tilde{\signal}$.

Through anonymisation, original speaker embeddings~$\xvo$ are substituted by pseudo-speaker embeddings $\xvp$, and used by the vocoder to reconstruct a speech signal using the F0 curve~$\fno$ and linguistic features~$\ppg$ extracted from the input speech signal corresponding to x-vector~$\xvo$. The new pseudo-speaker embedding will hence not \emph{match} any speaker-related information contained in~$\fno$ and~$\ppg$.  
This results in a \emph{mismatch} with the data distribution learned by the vocoder at training time.
It is our hypothesis that this mismatch is the source of vocoder drift.

We verified our hypothesis with an experiment in which we anonymised a set of utterances $\signal$ and computed  original x-vectors $\xvo$ and corresponding pseudo-speaker embeddings $\anon{\xvo} = \xvp$. Then, rather than synthesising new waveforms according to the usual approach $\vocoder{\fno}{\ppg}{\XVp}$, we compute instead $\vocoder{\fno}{\ppg}{\XVi}$, where $\xvi$ is an interpolation between $\xvo$ and $\xvp$:
\begin{equation}
\label{eq:interpolation}
    \xvi = \xvo + \lambda (\xvp - \xvo)
\end{equation}
The parameter $\lambda \in [0,1]$ acts to control the distance between $\xvi$ and either $\xvo$ or $\xvp$. 
In line with definitions presented in \secref{sec:drift_def}, we term $d(\xvo, \xvi)$ the \emph{target distance}.
The target distance can be interpreted to reflect the mismatch between the speaker embedding that would naturally complement $\fno$ and $\ppg$ and the embedding received by the vocoder.
By adjusting $\lambda$, we conducted a set of anonymisation experiments with increasing
target distances, i.e.\ higher values of~$\lambda$, equivalent to increasing feature mismatch.
For each experiment, we also measure the resulting vocoder drift.
A positive correlation between drift and target distance would then suggest that vocoder drift does indeed have some dependency on the mismatch between vocoder input features.

\subsection{Experiments and results}
We conducted experiments with values of ${\lambda=\{0, 1/3, 1/2, 1\}}$.
In the case of $\lambda=0$, \eqref{eq:interpolation} reduces to $\xvi = \xvo$, which corresponds to the absence of anonymisation (i.e.\ $\anon{\cdot}$ is not applied):
the system performs copy-synthesis.
Conversely, in the 
case of $\lambda=1$,
\eqref{eq:interpolation} reduces to $\xvi = \xvp$: the pseudo-speaker embedding is employed during synthesis as with usual anonymisation.
Values of $\lambda=1/3$ and $1/2$ correspond to different interpolations between $\xvo$ and $\xvp$.
We measured the target distance and vocoder drift for all four configurations.

Results are reported in \tabref{tab:drift}, which shows the target distance and drift in the first two columns of each set of results for each value of $\lambda$. Results are shown separately for LibriSpeech and VCTK datasets and for male and female subsets in both cases.
A degree of positive correlation between $\lambda$ and both the target distance and drift is apparent.
For $\lambda=0$, the target distance is always $0$ (since $\xvi=\xvo$) and the drift is consistently in the order of $0.28$.
For $\lambda=1/3$, the target distance increases to an average of $0.1$ and the drift to an average of $0.36$.
Both the target distance and drift increase further for higher values of $\lambda$:
such a correlation is evident when plotting the two metrics against one another for a whole data partition and different values of $\lambda$, as in \figref{fig:drift_trend}.
These results show that, the greater the degree of mismatch between input features, the greater is the vocoder drift.  This in turn implies that greater target distances incur less control over the x-vector space.
However, but not surprisingly, for copy-synthesis when $\lambda=0$, the drift is still substantial. 
For this configuration, there is no mismatch in the input features; those used for reconstruction are exactly those extracted from the input signal.
This suggests that a component of the drift stems from the intrinsic nature of the waveform reconstruction process.
In the following, we report an approach to compensate for the vocoder drift.

\begin{figure*}
    \centering
    \includegraphics[width=\linewidth]{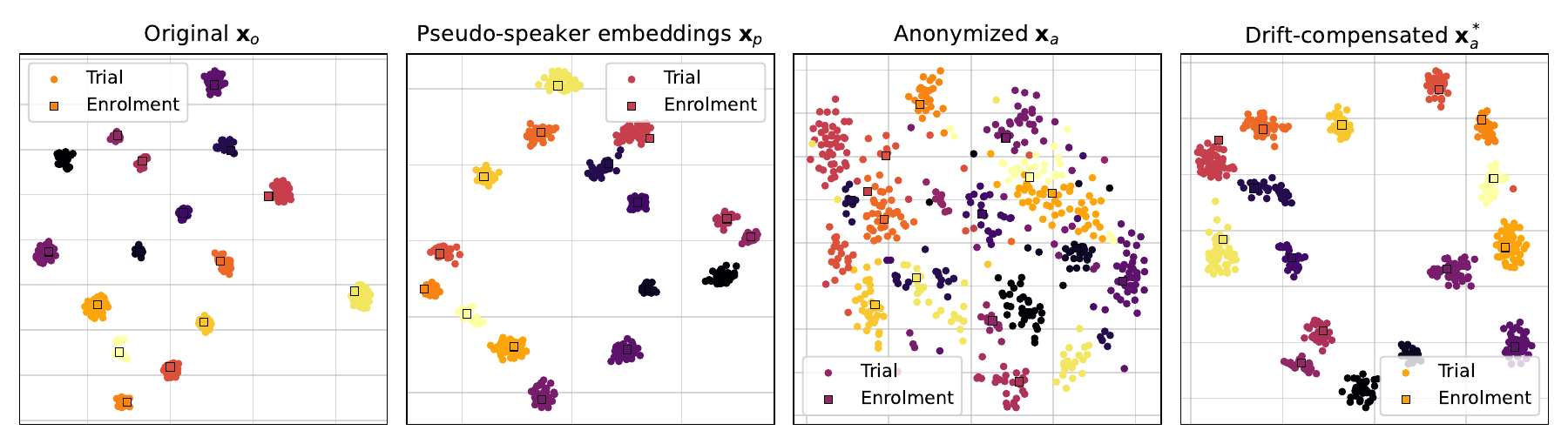}
    \caption{t-SNE visualisations of four different x-vector spaces and embeddings for the enrolment and trial utterances of the LibriSpeech dataset and female speakers. Different colours correspond to different speakers. From left to right: original x-vectors $\xvo$, pseudo-speaker embeddings $\xvp$, anonimised embeddings $\xva$, anonymised and drift-compensated embeddings $\xva^*$.}
    \label{fig:tsne_asv}
\end{figure*}

\section{Drift compensation}
\label{sec:align}
Vocoder drift, while advantageous in terms of anonymisation~\cite{drift}, can be  undesirable in that it prevents fine-grained control over the x-vector space.
Because the impact of the vocoder upon the x-vector space can dominate that of the anonymisation function, this lack of control impedes the design of better anonymisation functions. 
Hence, even if lower vocoder drift might initially degrade anonymisation performance, it may deliver better control over the x-vector space and then be beneficial to the future development of better anonymisation functions.
In this section, we introduce a new technique for vocoder drift compensation.  
It is based upon the iterative \emph{alignment} of $\xva$ to $\xvp$ at inference time.

\subsection{X-vector alignment}
Our goal is to adjust the matrix $\XVi$ so as to reduce the mismatch to $\ppg$ and $\fno$ in order then to reduce vocoder drift.
This adjustment can be formulated as an optimisation problem:
\begin{equation}
\label{eq:align}
    \XVi^* = \argmin_{\XVi} \text{   } d \Bigl(\,\, \underbrace{f\bigl(\vocoder{\fno}{\ppg}{\XVi}\bigr)}_{\xva} \,\, , \,\, \xvp \Bigr)
\end{equation}
where $d$ is again the cosine distance.
In essence, we seek to adjust $\XVi$ so as to minimise the cosine distance between $\xvp$ (the x-vector vocoder input) and $\xva$ (the x-vector extracted from its output).
The resulting, optimised matrix $\XVi^*$ is then used to synthesise an anonymised utterance $\Tilde{\signal}_a^*$, whose drift-compensated x-vector we denote as $\xva^*$.
We optimise the objective function directly at inference time via gradient descent.
With this approach, the drift can be arbitrarily reduced by any desired amount, at the cost of proportionately increasing the computation time required to synthesise the anonymised waveform. 

\subsection{Experiments and results}
We optimise~\eqref{eq:align}
at the utterance level
using Adam~\cite{adam} with a learning rate of $5\mathrm{e}{-3}$.
Optimisation runs for a maximum of 150 steps, but stops earlier if the drift falls below 0.05 (set arbitrarily to reduce processing time).
The impact of drift compensation is then observed by repeating the experiments described in Section~\ref{sec:causes}
but with $\XVi$ replaced by drift compensated versions $\XVi^*$ and by observing the reduction in vocoder drift.

Results are shown in the third columns of each block in \tabref{tab:drift}.
Drift compensation reduces the vocoder drift for all values of $\lambda$.
For $\lambda=\{0, 1/3\}$, 150 optimisation steps are generally sufficient for the drift to reach the lower bound of 0.05, 
for all datasets.
This is also the case for the LibriSpeech dataset for
$\lambda=\{1/2, 1\}$.  For the VCTK dataset, we obtain drift values of approximately 0.07 --- slightly higher than LibriSpeech, yet still considerably lower than the initial vocoder drift.
Informal listening tests show that drift compensation introduces no discernible
degradation to speech quality --- 
any differences are negligible to the point that signals generated with and without drift compensation are difficult to tell apart.

\subsection{Impact upon privacy protection}
If the vocoder drift is responsible for the bulk of anonymisation performance, and if drift compensation performs as intended, then the application of drift compensation is expected to result in degraded anonymisation performance.
We performed a set of ASV experiments to observe the impact. 
Experiments were conducted according to the protocol described in the VoicePrivacy Challenge 2022 evaluation plan~\cite{vp2022eval}. For each dataset, the experiment is run four times, each time using one of the set of x-vectors ($\xvo$, $\xvp$, $\xva$, $\xva^*$) for each utterance.
The results are reported in \tabref{tab:asv}.

As expected, low EERs for x-vectors $\xvo$ increase for $\xvp$ and even more noticeably for $\xva$, indicating the dominant impact of the vocoder upon anonymisation. 
This is especially evident in VCTK partitions, likely because of a domain mismatch with the HiFi-GAN vocoder which, in accordance with the VoicePrivacy 2022 protocol, is trained on \textit{LibriTTS-train-clean-100}.
EERs for x-vectors $\xva^*$ are close to those of $\xvp$, indicating successful vocoder drift compensation.
This result can also be observed visually in \figref{fig:tsne_asv}, which shows t-SNE visualisations~\cite{tsne} of all four x-vector embeddings for the LibriSpeech dataset and female speakers (both trial and enrolment utterances).  The effect of drift is clearly visible upon the comparison of the visualisations for $\xvp$ and $\xva$: in the latter, embeddings are notably more dispersed.  The visualisation for $\xva^*$ shows that drift compensation
reduces
the dispersion, giving compact clusters once more.

\begin{table}[t]
  \caption{ASV results (EER, \%) for VoicePrivacy 2022 test sets, using the same set of different x-vector speaker embeddings as in Figure~3.}
  \label{tab:asv}
\centering
\resizebox{\columnwidth}{!}{%
\begin{tblr}{lclccc}
\hline[1pt]
& $\xvo$ & $\xvp$ & $\xva$ & $\xva^*$ (comp.) \\ \hline
LibriSpeech (F) & $0.54$ & $2.51$ & $15.0$ & $2.75$ \\
LibriSpeech (M) & $0.88$ & $2.99$ & $14.5$ & $3.34$ \\
VCTK (F) & $1.13$ & $5.59$ & $25.3$ & $9.20$ \\
VCTK(M) & $0.17$ & $3.04$ & $18.5$ & $5.23$ \\ \hline[1pt]
\end{tblr}}
\end{table}

\section{Conclusions}
This paper shows that the mismatch between the representations of linguistic information, intonation and prosody and a substitute pseudo-speaker embedding is a source of vocoder drift -- the difference between a target x-vector and that which can be extracted from the synthesised output of popular approaches to speaker anonymisation.

While beneficial to anonymisation, vocoder drift can nonetheless be undesirable: it reduces fine-grained control over the x-vector space and hence impedes optimisation of the core anonymisation function.
Experiments show that a novel approach to compensate for vocoder drift through the iterative adjustment of pseudo-speaker embeddings to linguistic, intonation and prosodic components is effective in reducing the drift.
As expected, however, the loss of vocoder drift degrades anonymisation performance.
This result adds further weight to our previous findings that vocoder drift plays a substantial, but only superficial role in anonymisation; the vocoder drift can be learned and undone, or reversed by an adversary.

The anonymisation function remains to be of paramount importance since its impact cannot be, or is at least much more difficult to reverse. 
The design of better anonymisation functions should hence remain a focus in future work.
The alleviation of extraneous influences coming from vocoder drift delivers better control over the x-vector space and hence better potential to design more effective anonymisation functions in the future.
This does not preclude the study of disentangled representations or other vocoder schemes which might also offer complementary opportunities to reduce drift and improve control over the x-vector space.

\bibliographystyle{IEEEtran}
\balance
\bibliography{mybib}

\end{document}